\pdfoutput=1
\documentclass[preprint,showpacs,preprintnumbers,amsmath,amssymb,floatfix,endfloats*]{revtex4}

\usepackage{graphicx}
\usepackage{dcolumn}
\usepackage{bm}
\usepackage{amsfonts}

\DeclareMathAlphabet{\mathsfsl}{OT1}{cmr}{bx}{it}
\begin{document}
\title{Structural transformations during periodic deformation of low-porosity amorphous materials}
\author{Nikolai V. Priezjev$^{1,2}$ and Maxim A. Makeev$^{3}$}
\affiliation{$^{1}$Department of Mechanical and Materials
Engineering, Wright State University, Dayton, OH 45435}
\affiliation{$^{2}$National Research University Higher School of
Economics, Moscow 101000, Russia}
\affiliation{$^{3}$Department of Chemistry, University of
Missouri-Columbia, Columbia, MO 65211}
\date{\today}
\begin{abstract}

Atomistic simulations are employed to study structural evolution of
pore ensembles in binary glasses under periodic shear deformation
with varied amplitude. The consideration is given to porous systems
in the limit of low porosity. The initial ensembles of pores are
comprised of multiple pores with small sizes, which are
approximately normally distributed. As periodic loading proceeds,
the ensembles evolve into configurations with a few large-scale
pores and significantly reduced number of small pores. These
structural changes are reflected in the skewed shapes of the
pore-size distribution functions and the appearance of a distinct
peak at large length scales after hundreds of shear cycles.
Moreover, periodic shear causes substantial densification of solid
domains in the porous systems. The structural evolution of pore
ensembles is found to stem from the formation of shear-band like
regions of enhanced particle mobility after a number of transient
cycles. The spatial extent of increased mobility depends strongly on
the strain amplitude. A scaling theory is developed to qualitatively
describe the transformation of the pore initial configurations of
small-size voids into larger-scale void agglomerates.

\end{abstract}



\maketitle

\section{Introduction}

The design of the optimal microstructural architecture for metallic
glasses with enhanced ductility is important for various structural
applications~\cite{SchroersNat13}. It is well recognized by now that
homogeneous, pore-free metallic glasses yield via the formation of
narrow shear bands leading to fracture and breaking of the material.
However, the tensile ductility can be improved by spatially
constraining shear bands and introducing porous heterostructures in
an amorphous solid~\cite{SchroersNat13}. Thus, it was recently shown
experimentally and numerically that a regular array of pores inside
or at the surface of bulk metallic glasses changes the stress field
upon deformation and guide the formation of shear bands along the
domains with multiple pores~\cite{Bargmann14,GaoSci16,Song17,Luo18}.
More recently, it was demonstrated that the strength of nanoporous
metallic glasses depends sensitively on the porosity and pore shape,
and the maximum strength is attained by localizing shear bands for
sufficiently low values of porosity~\cite{ZhouCMS18}. Using
atomistic simulations, it was found that elastic moduli follow a
power-law increase as a function of the average glass density in
amorphous solids with random porous structures~\cite{Priezjev17s,
Priezjev18t,Priezjev18tp,Priezjev18c}.  With further increasing
strain, the porous structure deforms significantly and nearby pores
tend to coalesce with each other leading to formation of a dominant
cavity at high strain~\cite{Priezjev17s,Priezjev18t,Priezjev18tp,
Priezjev18c}. Nevertheless, the evolution of the pore size
distribution and formation of shear bands in periodically deformed
glasses still need to be thoroughly explored.

\vskip 0.05in

In the past few years, large-scale molecular dynamics simulations
were extensively used to investigate the dynamic response of
homogeneous amorphous materials to oscillatory shear
deformation~\cite{Priezjev13,Sastry13,Reichhardt13,Priezjev14,
IdoNature15,Priezjev16,Kawasaki16,Yang16,Priezjev16a,Sastry17,
Priezjev17,Hecke17,Keblinsk17,Priezjev18,Priezjev18a,NVP18strload,
SastryBands18}.  It was found that the relaxation dynamics at small
strain amplitudes depends strongly on the preparation history. In
particular, it was shown that after a few training cycles, slowly
annealed glasses start to deform reversibly and the trajectory of
each atom repeats itself at zero temperature~\cite{Reichhardt13,
Sastry13,IdoNature15,Priezjev16,Priezjev16a}. Interestingly, during
periodic deformation below yield, particles with relatively large
displacements form clusters, which, depending on the strain
amplitude, can be comparable to the system size~\cite{IdoNature15}.
On the other hand, poorly annealed glasses were found to relocate
toward progressively lower potential energy levels when subjected to
periodic loading in the elastic range~\cite{Sastry13,Sastry17,
Priezjev18,Priezjev18a,NVP18strload}. The yielding transition
typically occurs after a certain number of cycles, and it is
accompanied by the formation of a system-spanning shear band in
sufficiently large systems and, as a result, higher potential energy
levels~\cite{Priezjev17,Sastry17,Priezjev18a,SastryBands18}.
Although some of these features, including the decay of the
potential energy and transient stress response, were recently
detected in highly porous binary glasses under periodic
shear~\cite{PriezMak18cyc}, the exact mechanisms of the pore and
glass phase redistribution as well as the nature of the yielding
transition remain not fully understood.

\vskip 0.05in

In this paper, we use molecular dynamics simulations to investigate
the effect of cyclic loading on evolution of porous structures in
low-porosity binary glasses. We find that under periodic
deformation, the initial ensemble of pores, which are approximately
normally distributed, gradually evolve into configurations with a
few large-scale pores that are energetically favorable. The
structural transformations of the pore and glass phases are
quantified via the pore-size and local density distribution
functions. The results are rationalized by estimating the stability
and lifetime of a pore, which depend on the pore size, surface
energy, and strain-driven diffusion of atoms near the pore.

\vskip 0.05in

The paper is structured as follows. The details of molecular
dynamics simulations including parameter values, interaction
potentials, preparation and deformation protocols are provided in
the next section.   The time dependence of the potential energy
series, variation of shear stress and analysis of pore size and
local density distribution functions are presented in
section\,\ref{sec:Results}. The summary and concluding remarks are
given in the last section.

\section{Details of MD simulations}
\label{sec:MD_Model}


In the present study, the molecular dynamics simulations were
carried out on a model glass, which is represented by a binary
(80:20) mixture first introduced by Kob and Andersen
(KA)~\cite{KobAnd95}.  In the KA model, any two atoms of types
$\alpha,\beta=A,B$ interact via the Lennard-Jones (LJ) potential of
the form:
\begin{equation}
V_{\alpha\beta}(r)=4\,\varepsilon_{\alpha\beta}\,\Big[\Big(\frac{\sigma_{\alpha\beta}}{r}\Big)^{12}\!-
\Big(\frac{\sigma_{\alpha\beta}}{r}\Big)^{6}\,\Big],
\label{Eq:LJ_KA}
\end{equation}
with the following parametrization $\varepsilon_{AA}=1.0$,
$\varepsilon_{AB}=1.5$, $\varepsilon_{BB}=0.5$, $\sigma_{AB}=0.8$,
and $\sigma_{BB}=0.88$, and $m_{A}=m_{B}$~\cite{KobAnd95}. The LJ
potential is truncated at the cutoff radius
$r_{c,\,\alpha\beta}=2.5\,\sigma_{\alpha\beta}$.  As usual, we
express physical quantities in the reduced LJ units of length, mass,
energy, and time; namely, $\sigma=\sigma_{AA}$, $m=m_{A}$,
$\varepsilon=\varepsilon_{AA}$, and
$\tau=\sigma\sqrt{m/\varepsilon}$. The Newton's equations of motion
for each atom were solved using the velocity-Verlet
scheme~\cite{Allen87} with the integration time step $\triangle
t_{MD}=0.005\,\tau$.

\vskip 0.05in


In order to obtain porous samples, we follow the preparation
procedure introduced in the recent MD studies~\cite{Kob11,Kob14}.
First, the binary mixture of $300\,000$ particles was thoroughly
equilibrated at constant volume and at the temperature of
$1.5\,\varepsilon/k_B$. Here, $k_B$ stands for the Boltzmann
constant.  The glass transition temperature of the KA model is
$T_g\approx0.435\,\varepsilon/k_B$~\cite{KobAnd95}.  Second, the
system was instantaneously quenched across the glass transition and
allowed to evolve during the time interval of $10^{4}\,\tau$ at
constant volume and temperature $T_{LJ}=0.05\,\varepsilon/k_B$. As a
result of the coarsening process and material solidification, the
porous structure is developed in samples with the average glass
density $\rho\sigma^{3}=0.9$.

\vskip 0.05in


After the porous structure was formed, the glass was subjected to
periodic shear deformation at constant volume as follows:
\begin{equation}
\gamma(t)=\gamma_{0}\,\,\textrm{sin}(2\pi t / T),
\label{Eq:strain}
\end{equation}
where $\gamma(t)$ is time-dependent shear strain, $\gamma_{0}$ is
the strain amplitude, $0 \leqslant \gamma_{0} \leqslant 0.12$, and
$T$ is the oscillation period. In what follows, the period is fixed
to $T=500\,\tau$, and the simulations were performed during $500$
cycles for each value of the strain amplitude. To avoid ambiguity,
we denote temperature by $T_{LJ}$, while the oscillation period is
indicated by $T$.  During periodic shear, the temperature
$T_{LJ}=0.05\,\varepsilon/k_B$ was maintained via the
Nos\'{e}-Hoover thermostat~\cite{Lammps}.  In addition, the
so-called Lees-Edwards periodic boundary conditions were imposed
along the plane of shear~\cite{Allen87}. The MD simulations were
performed only for one realization of disorder due to relatively
large system size, which is required to avoid finite size
effects~\cite{Kob11,Kob14}.   During production runs, several
characteristics including potential energy, shear stress, and
temperature, as well as positions of all atoms were periodically
saved for the postprocessing analysis.

\section{Results}
\label{sec:Results}


Before presenting our main results, we briefly revisit the dynamics
of the coarsening process that leads to the formation of porous
glassy media. As discussed in the recent MD
studies~\cite{Kob11,Kob14,Priezjev17s,Makeev18,Priezjev18t,Priezjev18c,
Priezjev18tp,PriezMak18cyc}, the binary mixture was first
instantaneously quenched across the glass transition and then
allowed to evolve freely at constant volume and
$T_{LJ}=0.05\,\varepsilon/k_B$ during the time period of
$10^{4}\,\tau$. It was previously demonstrated that this time
interval is sufficiently long, such that the typical size of solid
domains crosses over to logarithmically slow
growth~\cite{Kob11,Kob14}. It will be shown below, however, that in
the absence of mechanical deformation, the porous structure
undergoes a noticeable change during the subsequent time interval of
$2.5\times10^5\tau$.   It should be pointed out that the essential
feature of the coarsening process is the constraint of constant
volume, which leads to built in tensile stresses and negative
pressures~\cite{Kob11,Kob14,Priezjev17s,Makeev18}. Thus, the average
pressure was estimated to be $P\approx-0.73\,\varepsilon/\sigma^3$
for porous samples with the average glass density
$\rho\sigma^{3}=0.9$ and temperature
$T_{LJ}=0.05\,\varepsilon/k_B$~\cite{Makeev18}.

\vskip 0.05in

\begin{figure}[t]
\includegraphics[width=12.cm,angle=0]{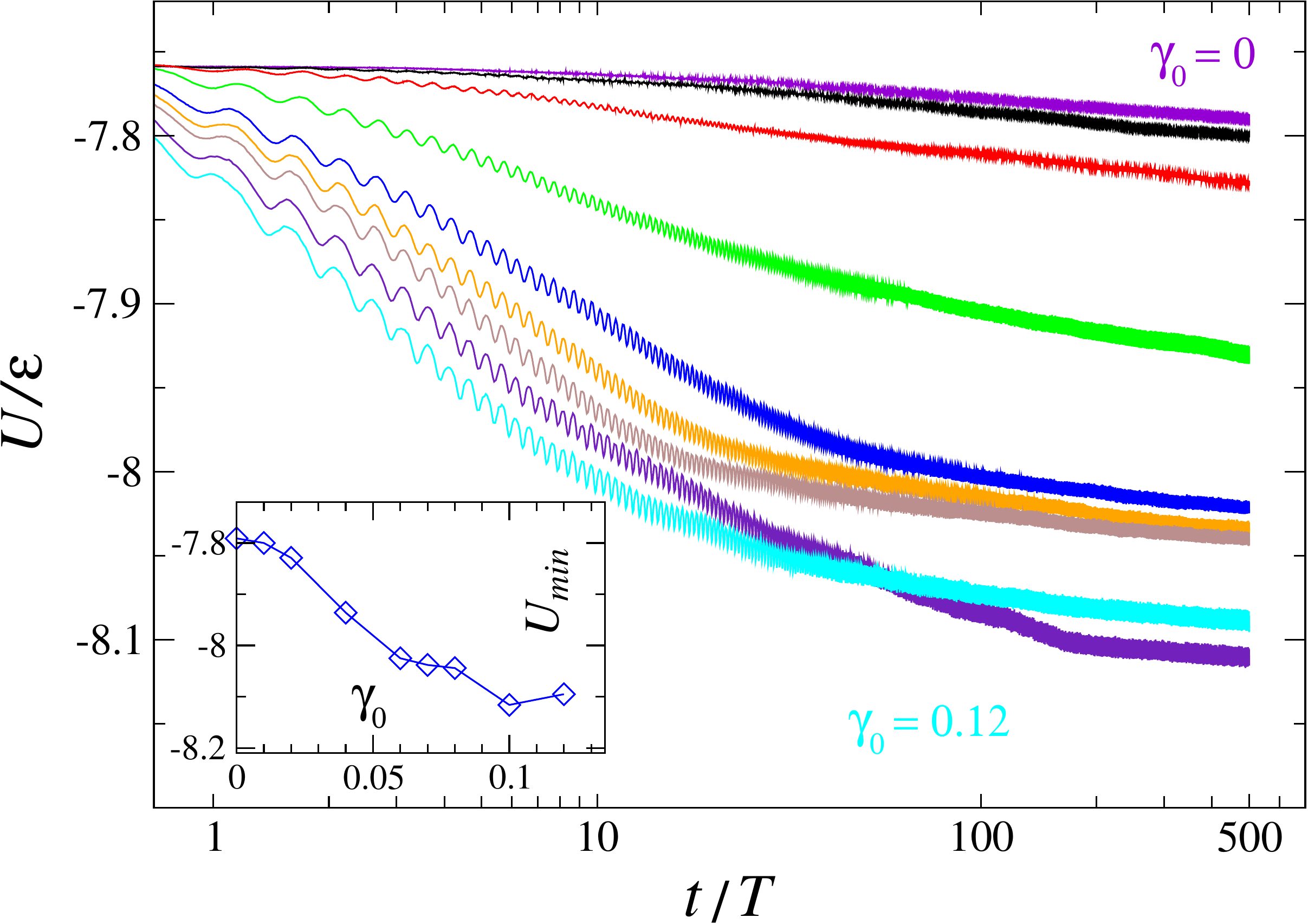}
\caption{(Color online) The potential energy, $U$, as a function of
time for strain amplitudes $\gamma_0=0$, $0.01$, $0.02$, $0.04$,
$0.06$, $0.07$, $0.08$, $0.10$, and $0.12$ (from top to bottom). The
average glass density is $\rho\sigma^{3}=0.9$ and the oscillation
period is $T=500\,\tau$.  The inset shows the variation of $U$ after
500 cycles versus the strain amplitude.}
\label{fig:poten_cycle_number}
\end{figure}

The time dependence of the potential energy per atom for the
indicated strain amplitudes is reported in
Fig.\,\ref{fig:poten_cycle_number} during 500 cycles. In each case,
the potential energy continues to decrease over consecutive cycles,
and, except for $\gamma_0 = 0.10$ and $0.12$, the minimum of the
potential energy after 500 cycles becomes deeper with increasing
strain amplitude. This trend is associated with significant
rearrangement of the glass phase in the driven porous systems. Note
that in the quiescent sample ($\gamma_0 = 0$), the decrease in the
potential energy is less pronounced as it corresponds to the aging
process at constant volume during the time interval
$500\,T=2.5\times10^5\tau$. Somewhat surprisingly, we observed that
the potential energy for cyclic loading with $\gamma_0 = 0.10$ is
lower than for $\gamma_0 = 0.12$. We attribute this behavior to the
particular realization of disorder, which, upon cycling, resulted in
larger pore structures and denser glass phase when $\gamma_0 = 0.10$
(discussed below).  It should be mentioned that for all strain
amplitudes, the potential energy levels are deeper for samples with
the average glass density $\rho\sigma^{3}=0.9$ than for
$\rho\sigma^{3}=0.5$ reported previously~\cite{PriezMak18cyc}, even
though only 500 cycles were applied in the former case and 2000
cycles in the latter case.

\vskip 0.05in

\begin{figure}[t]
\includegraphics[width=12.cm,angle=0]{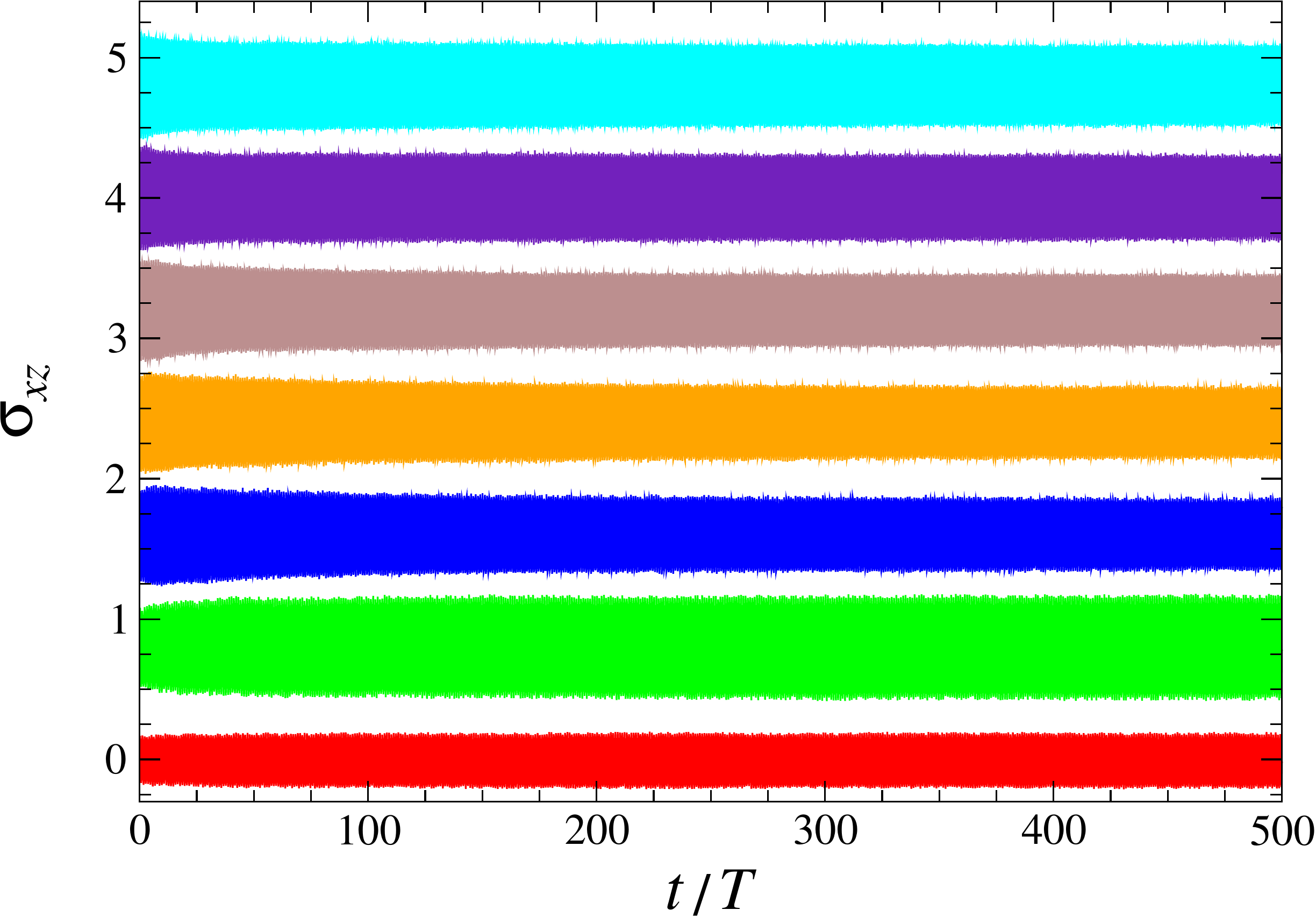}
\caption{(Color online) The time-dependent shear stress
$\sigma_{xz}$ (in units of $\varepsilon\sigma^{-3}$) for strain
amplitudes $\gamma_0=0.02$, $0.04$, $0.06$, $0.07$, $0.08$, $0.10$,
and $0.12$ (from bottom to top). The data are displaced upward by
$0.8\,\varepsilon\sigma^{-3}$ for $\gamma_0=0.04$, by
$1.6\,\varepsilon\sigma^{-3}$ for $\gamma_0=0.06$, by
$2.4\,\varepsilon\sigma^{-3}$ for $\gamma_0=0.07$, by
$3.2\,\varepsilon\sigma^{-3}$ for $\gamma_0=0.08$, by
$4.0\,\varepsilon\sigma^{-3}$ for $\gamma_0=0.10$, and by
$4.8\,\varepsilon\sigma^{-3}$ for $\gamma_0=0.12$.  }
\label{fig:stress_cycle_500}
\end{figure}

The variation of shear stress, $\sigma_{xz}$, as a function of time
during 500 cycles is presented in Fig.\,\ref{fig:stress_cycle_500}
for different strain amplitudes.   It can be seen that following
about 50 transient cycles, the stress amplitude remains nearly
constant.  The relative trends, however, are distinctly different
for the cases $\gamma_0 \leqslant 0.04$ and $\gamma_0 \geqslant
0.06$ during the first 50 cycles.   At small strain amplitudes,
$\gamma_0 \leqslant 0.04$, the stress amplitude gradually increases
over the first 50 cycles, due to a rapid decrease of the potential
energy, and then it saturates at a constant value corresponding to a
nearly reversible, elastic deformation. A similar effect was
observed for rapidly quenched, homogeneous binary glasses subjected
to oscillatory shear deformation with strain amplitudes below
yield~\cite{Priezjev18,Priezjev18a}.  In contrast, the amplitude of
stress oscillations at $\gamma_0 \geqslant 0.06$ decreases after a
few cycles and becomes smaller than for the case $\gamma_0 = 0.04$,
indicating large-scale plastic deformation. These results are
qualitatively similar to the behavior of shear stress for lower
glass density samples $\rho\sigma^{3}=0.5$ reported in the previous
study~\cite{PriezMak18cyc}, although the magnitude of stress
variations are significantly larger for the case
$\rho\sigma^{3}=0.9$ presented herein.

\vskip 0.05in

\begin{figure}[t]
\includegraphics[width=15.cm,angle=0]{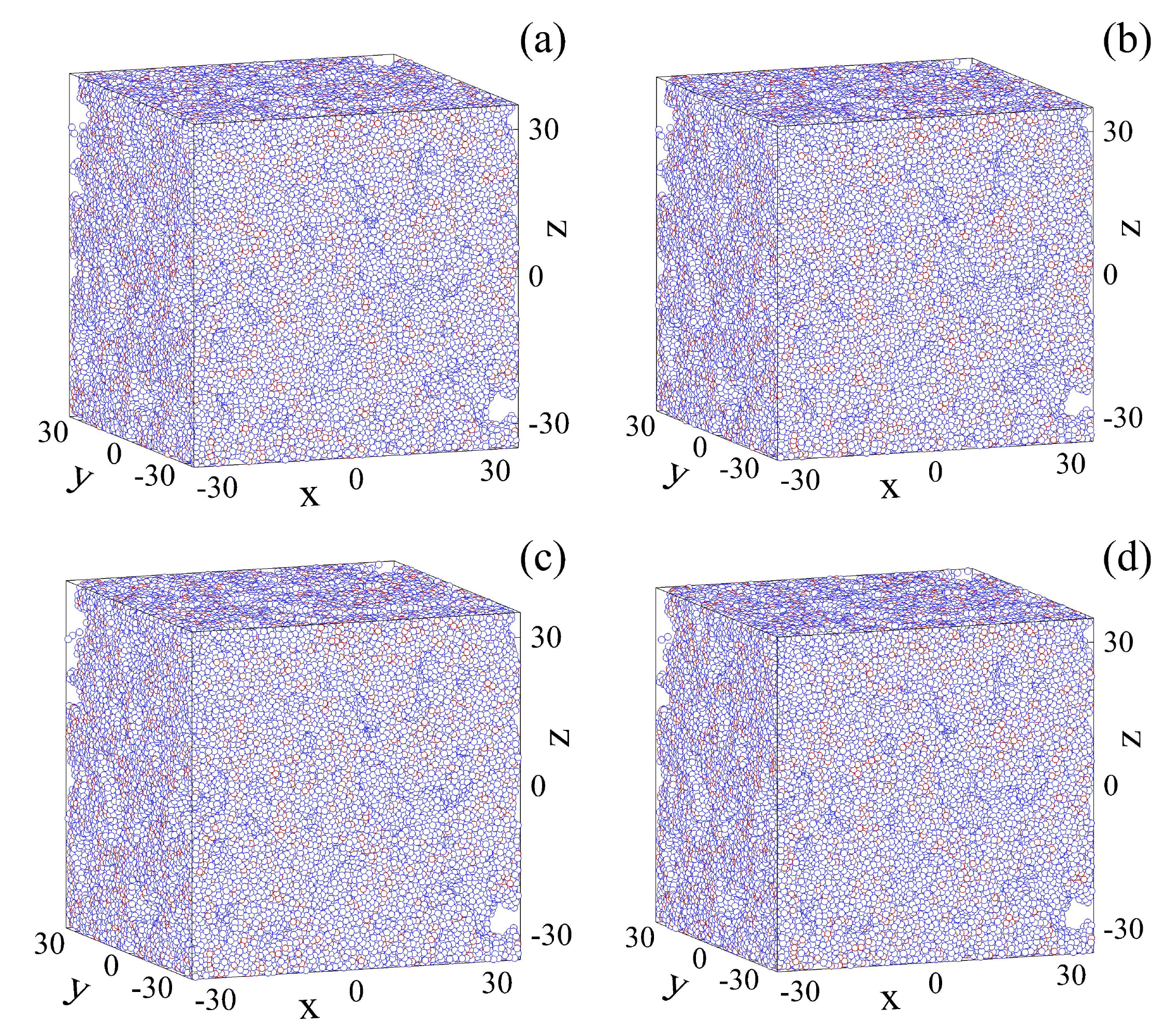}
\caption{(Color online) Consecutive snapshots of the quiescent glass
($\gamma_0=0$) after the time intervals (a) $T$, (b) $10\,T$, (c)
$100\,T$, and (d) $500\,T$.  The oscillation period is $T=500\,\tau$
and the average glass density is $\rho\sigma^{3}=0.9$. }
\label{fig:snapshot_gamma0_00}
\end{figure}
\begin{figure}[t]
\includegraphics[width=15.cm,angle=0]{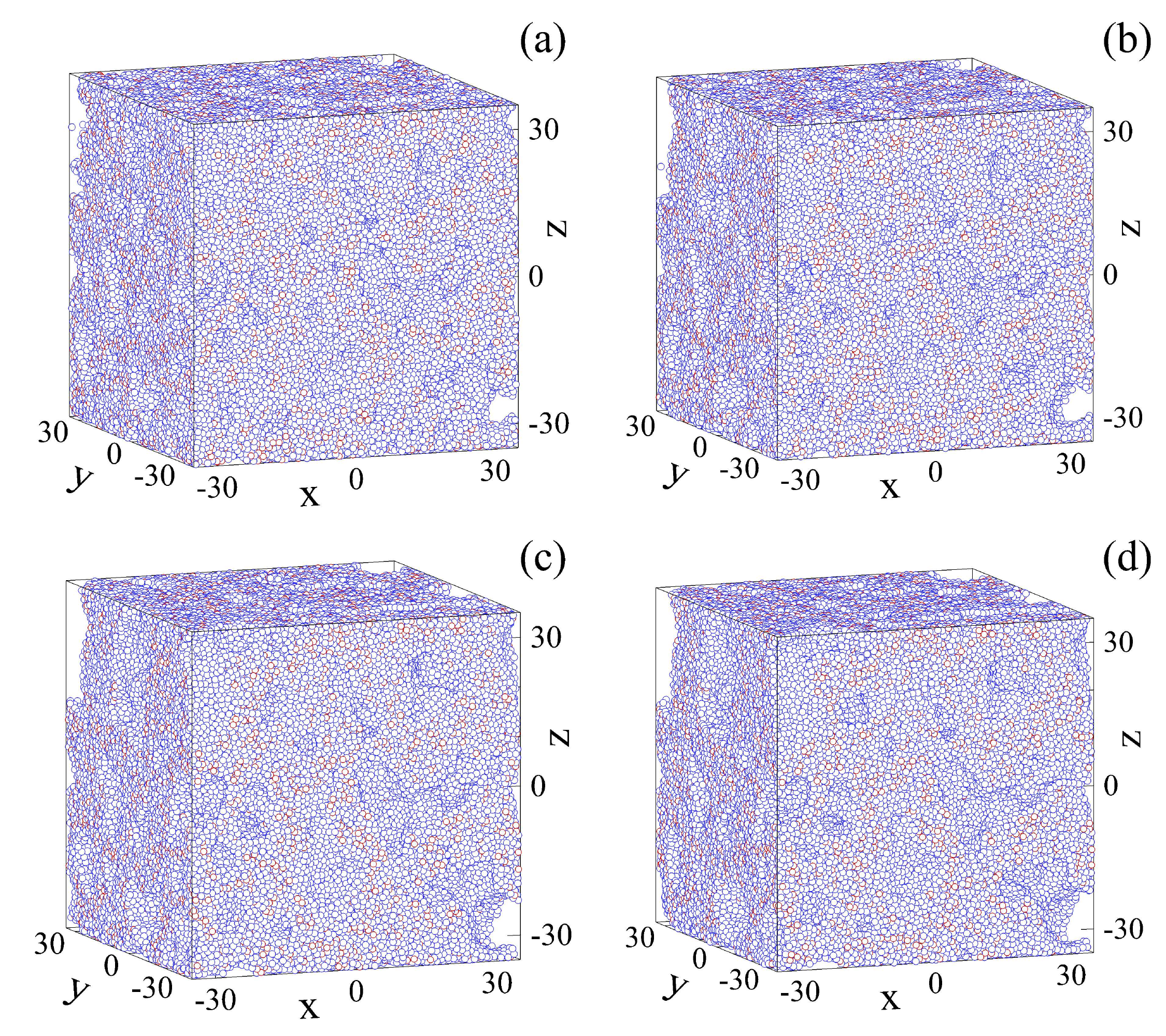}
\caption{(Color online) System snapshots during periodic deformation
with the strain amplitude $\gamma_0=0.04$ after (a) 1-st, (b) 10-th,
(c) 100-th, and (d) 500-th cycle at zero strain.}
\label{fig:snapshot_gamma0_04}
\end{figure}
\begin{figure}[t]
\includegraphics[width=15.cm,angle=0]{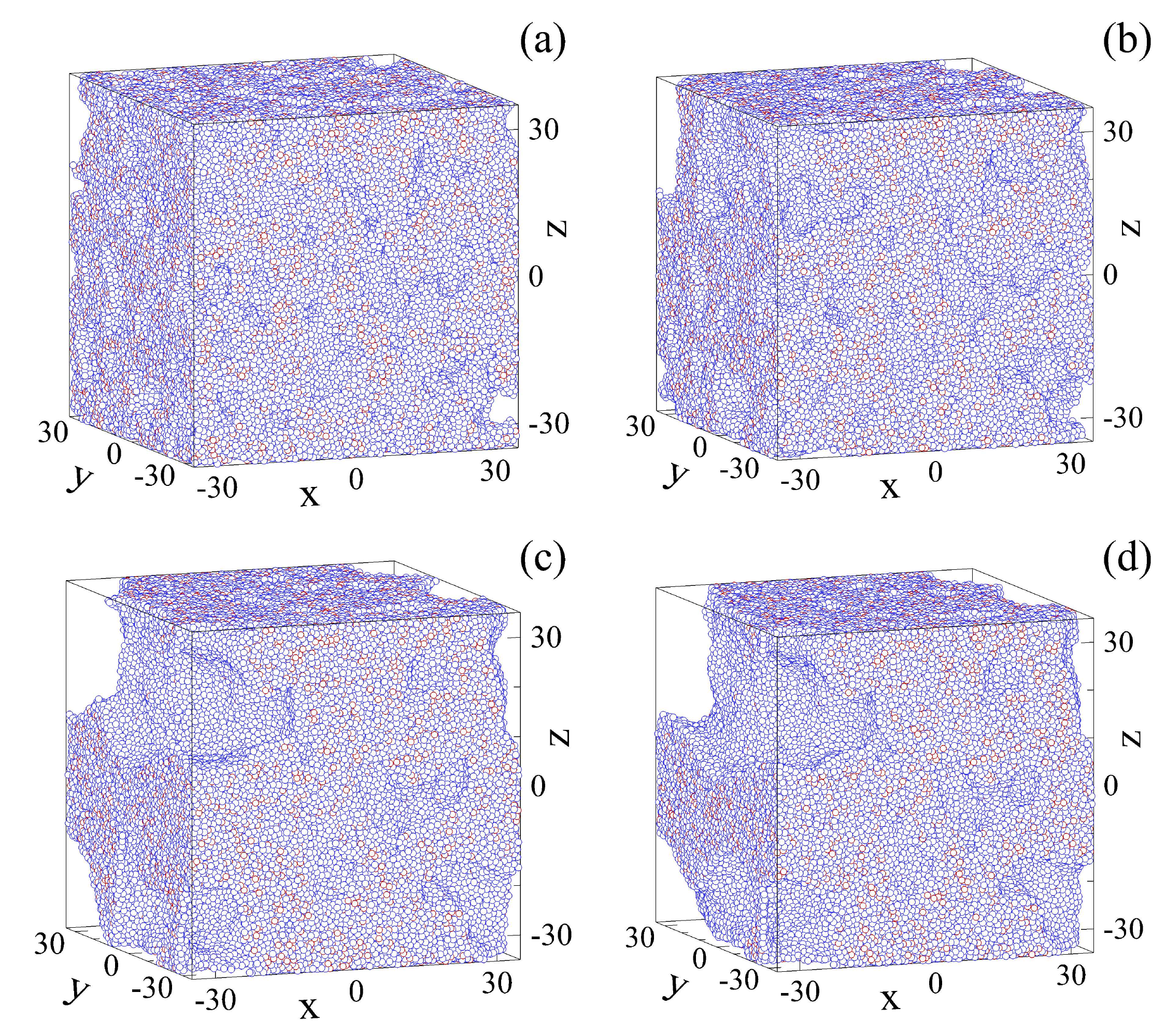}
\caption{(Color online) Spatial configurations of atoms during
oscillatory shear deformation with the strain amplitude
$\gamma_0=0.06$ after (a) 1-st, (b) 10-th, (c) 100-th, and (d)
500-th cycle.}
\label{fig:snapshot_gamma0_06}
\end{figure}
\begin{figure}[t]
\includegraphics[width=15.cm,angle=0]{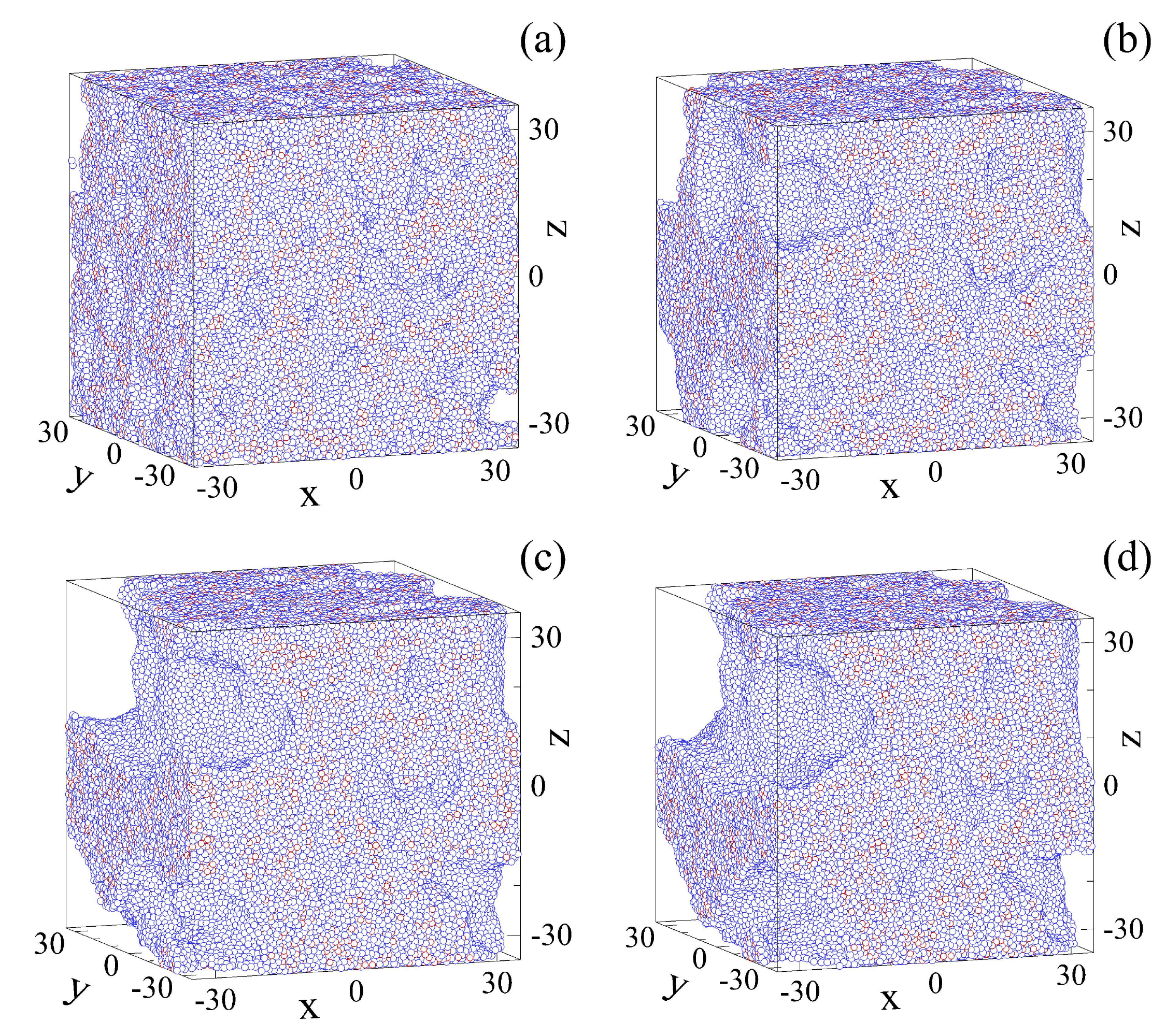}
\caption{(Color online) Atom positions at zero strain during
periodic loading with the strain amplitude $\gamma_0=0.08$ after (a)
1-st, (b) 10-th, (c) 100-th, and (d) 500-th cycle.}
\label{fig:snapshot_gamma0_08}
\end{figure}
\begin{figure}[t]
\includegraphics[width=15.cm,angle=0]{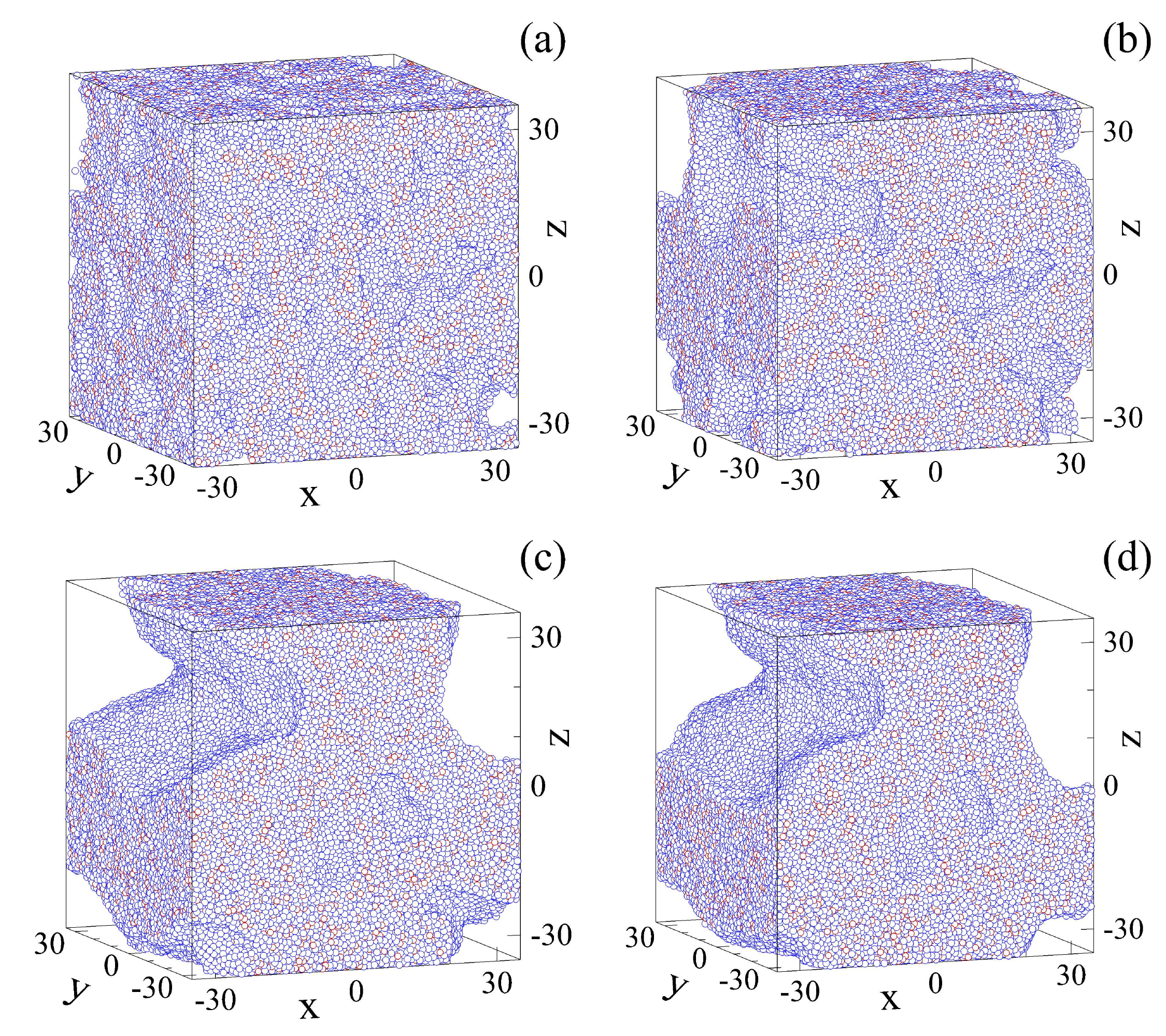}
\caption{(Color online) Instantaneous snapshots of the porous glass
during cyclic loading with the strain amplitude $\gamma_0=0.10$
after (a) 1-st, (b) 10-th, (c) 100-th, and (d) 500-th cycle.}
\label{fig:snapshot_gamma0_10}
\end{figure}
\begin{figure}[t]
\includegraphics[width=15.cm,angle=0]{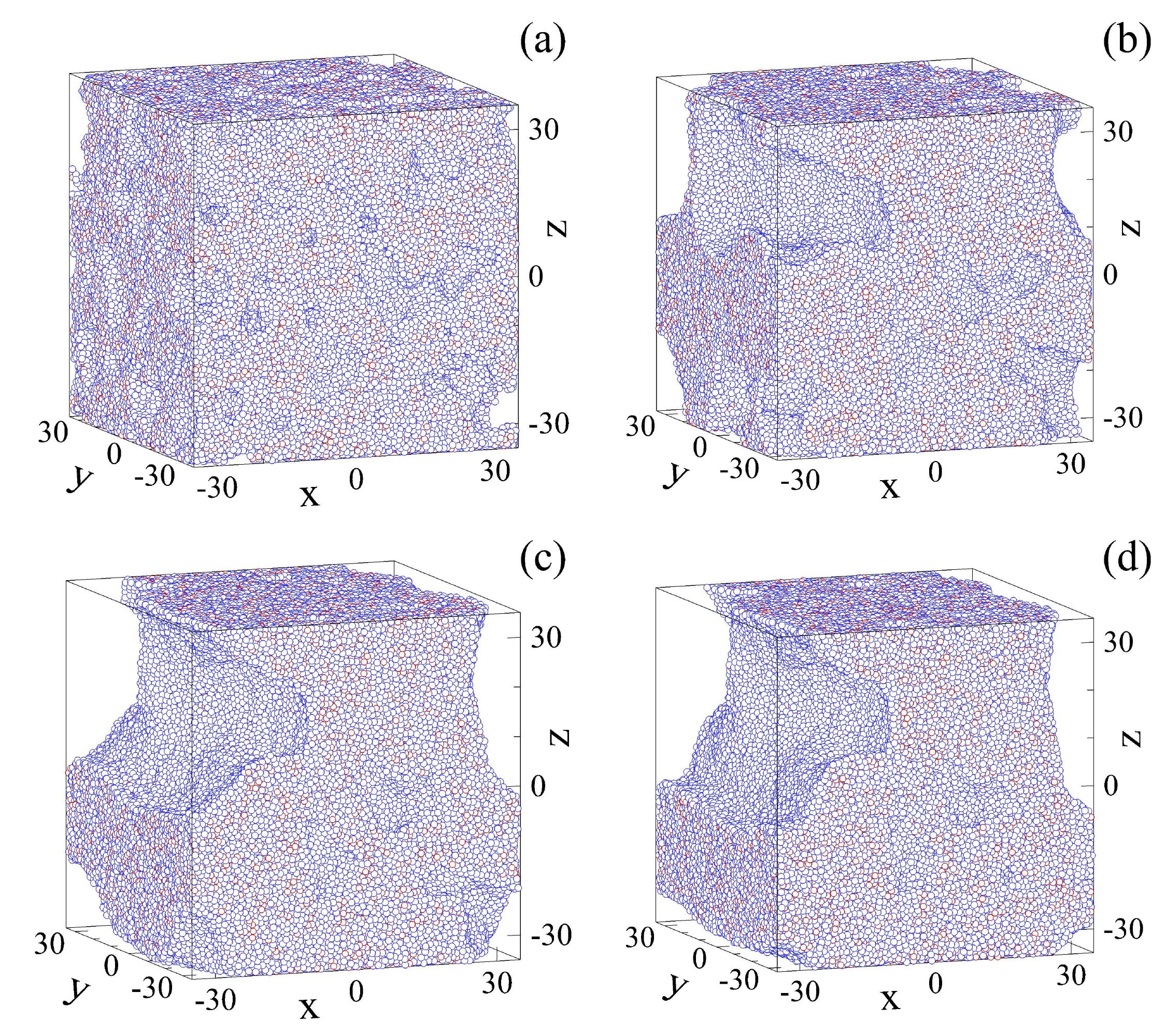}
\caption{(Color online) Atomic configurations during cyclic shear
deformation with the strain amplitude $\gamma_0=0.12$ after (a)
1-st, (b) 10-th, (c) 100-th, and (d) 500-th cycle.}
\label{fig:snapshot_gamma0_12}
\end{figure}
\begin{figure}[t]
\includegraphics[width=15.cm,angle=0]{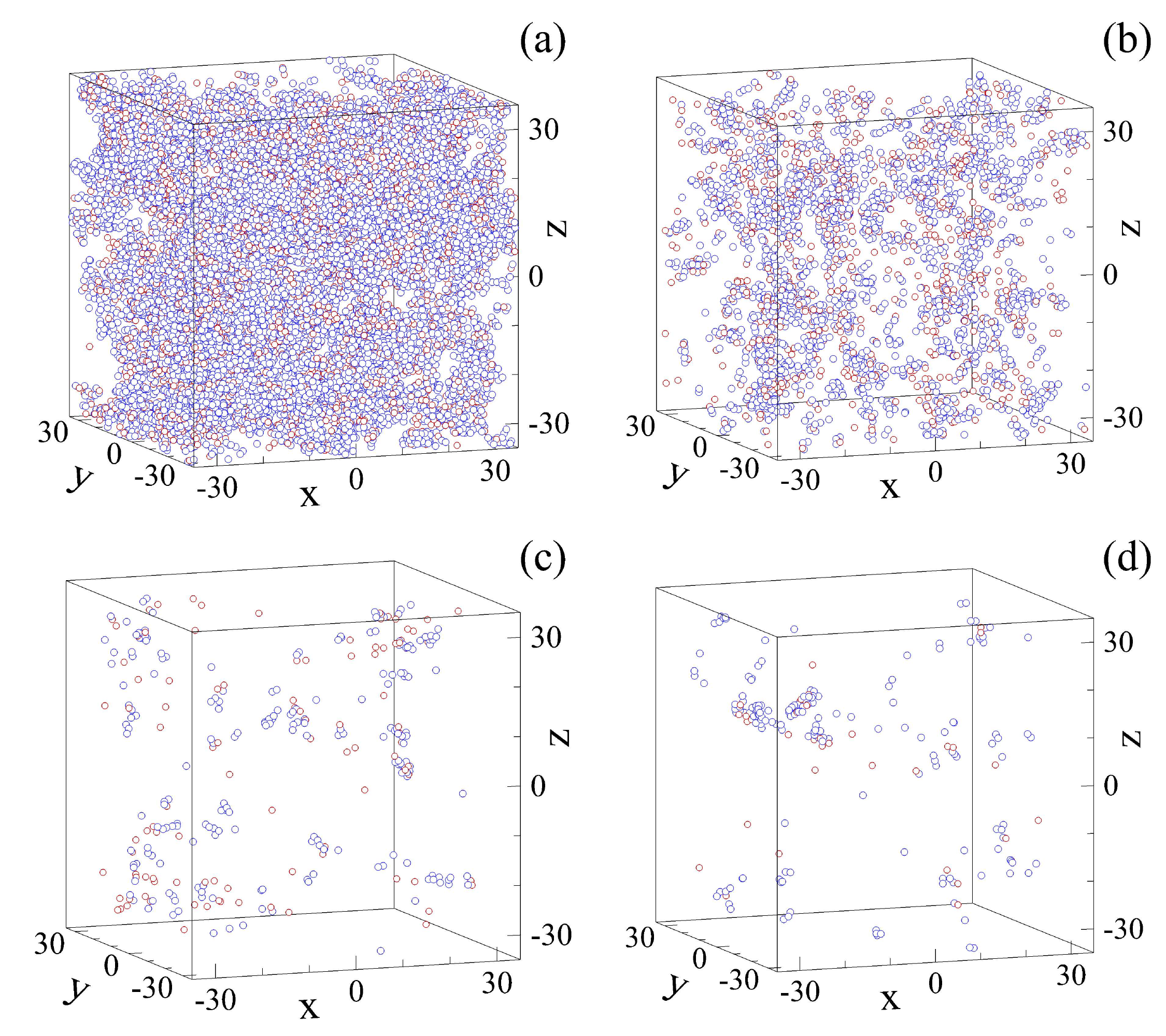}
\caption{(Color online) The positions of atoms whose displacements
during one cycle is greater than $0.6\,\sigma$ for the cycle numbers
(a) 1, (b) 10, (c) 100, and (d) 500.  The strain amplitude is
$\gamma_0=0.04$.  The same data as in
Fig.\,\ref{fig:snapshot_gamma0_04}. }
\label{fig:collective_gamma0_04}
\end{figure}

A series of instantaneous snapshots of the porous glasses for
selected strain amplitudes is illustrated in
Figs.\,\ref{fig:snapshot_gamma0_00}--\ref{fig:snapshot_gamma0_12}.
In all panels, the snapshots are taken after the indicated number of
cycles at zero strain.   It can be observed that the porous
structure remains nearly unchanged during 500 cycles for the
quiescent sample shown in Fig.\,\ref{fig:snapshot_gamma0_00} and for
the strain amplitude below the yield point, $\gamma_0=0.04$, see
Fig.\,\ref{fig:snapshot_gamma0_04}.   While local plastic events are
strongly suppressed in the quiescent glass at the low temperature
$T_{LJ}=0.05\,\varepsilon/k_B$ (not shown), the collective
irreversible displacements of atoms become abundant during the first
several cycles at $\gamma_0=0.04$; however, they decay quickly  over
consecutive cycles, indicating nearly reversible shear deformation
(see Fig.\,\ref{fig:collective_gamma0_04}).   This behavior is
similar to the decrease in the volume occupied by atoms with large
nonaffine displacements in quickly annealed, homogeneous binary
glasses subjected to repetitive subyield
cycling~\cite{Priezjev18,Priezjev18a}.

\vskip 0.05in

\begin{figure}[t]
\includegraphics[width=15.cm,angle=0]{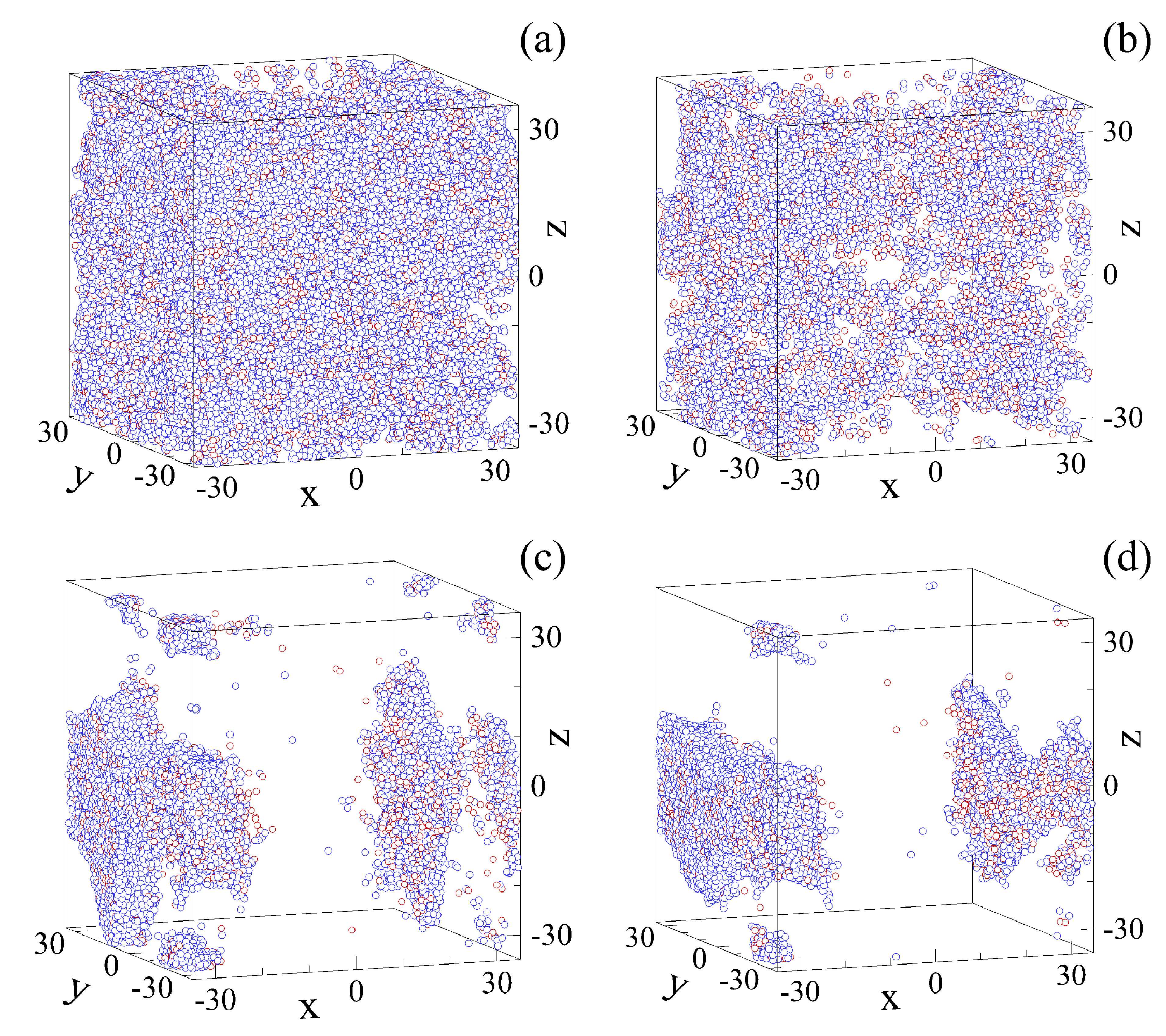}
\caption{(Color online) Configurations of atoms with displacements
during one back-and-forth cycle greater than $0.6\,\sigma$ for the
cycle numbers (a) 1, (b) 10, (c) 100, and (d) 500.  The strain
amplitude is $\gamma_0=0.06$. The same data as in
Fig.\,\ref{fig:snapshot_gamma0_06}. }
\label{fig:collective_gamma0_06}
\end{figure}
\begin{figure}[t]
\includegraphics[width=15.cm,angle=0]{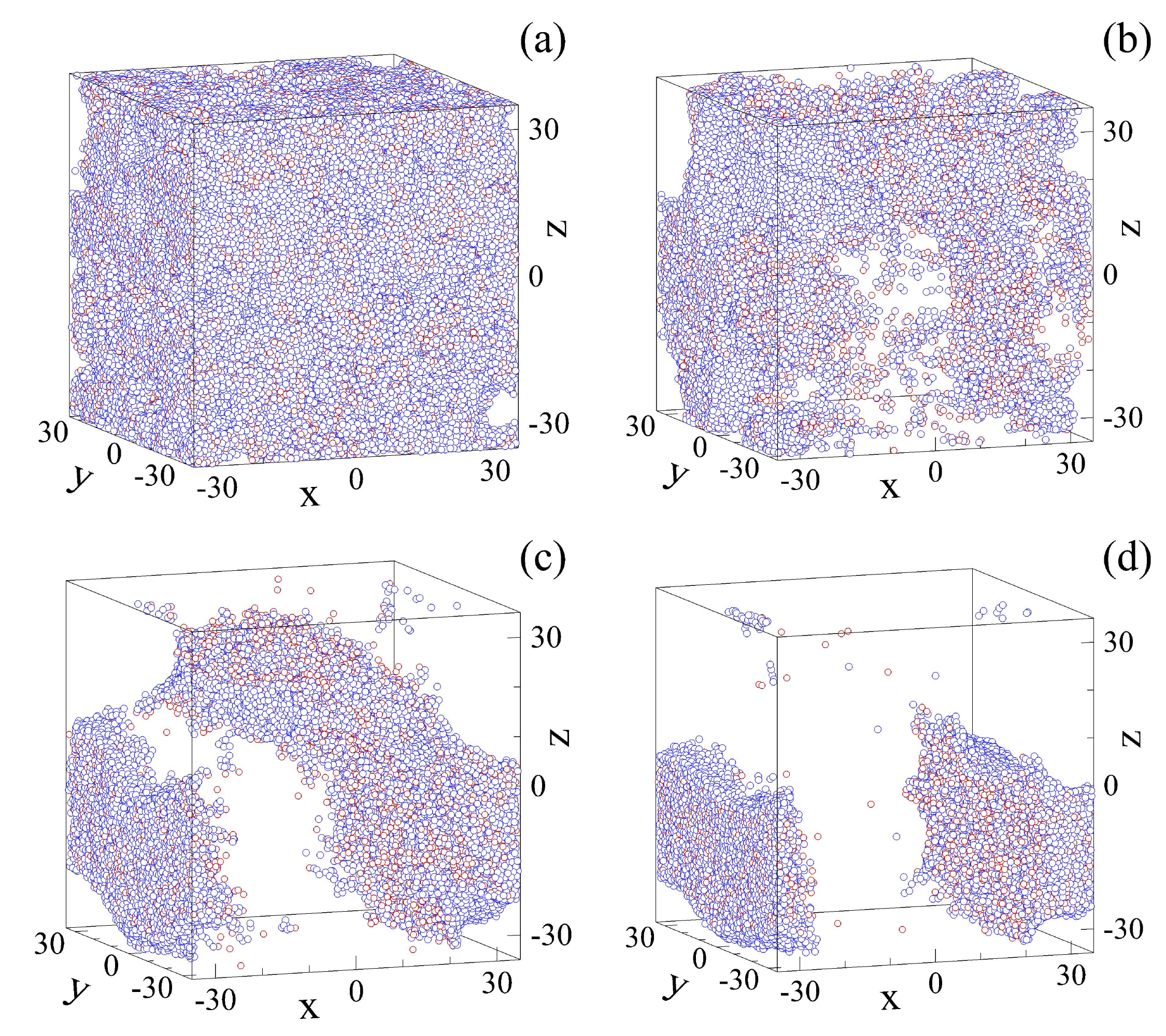}
\caption{(Color online) The positions of mobile atoms with
displacements during one cycle greater than $0.6\,\sigma$ for the
cycle numbers (a) 1, (b) 10, (c) 100, and (d) 500. The strain
amplitude is $\gamma_0=0.10$.  The same data as in
Fig.\,\ref{fig:snapshot_gamma0_10}. }
\label{fig:collective_gamma0_10}
\end{figure}

In sharp contrast, periodic loading with large strain amplitudes,
$\gamma_0 \geqslant 0.06$, results in significant redistribution of
pores and formation of a dominant cavity after 500 cycles, see
Figs.\,\ref{fig:snapshot_gamma0_06}--\ref{fig:snapshot_gamma0_12}.
As shown in Figs.\,\ref{fig:collective_gamma0_06} and
\ref{fig:collective_gamma0_10}, most of the atoms undergo large
irreversible displacements during the first few transient cycles,
followed by formation of a permanent shear band.  Thus, the
migration and coalescence of pores is enhanced in the regions
populated with mobile atoms.   It can be seen in
Figs.\,\ref{fig:snapshot_gamma0_06} and \ref{fig:snapshot_gamma0_10}
that pores are absent after 100 cycles inside the shear bands shown
in Figs.\,\ref{fig:collective_gamma0_06} and
\ref{fig:collective_gamma0_10}.   Note that similar trends are
evident for other strain amplitudes, although they are not reported
here for brevity.   We also comment that the orientation of shear
bands in Figs.\,\ref{fig:collective_gamma0_06} and
\ref{fig:collective_gamma0_10} is perpendicular to the plane of
shear.    Such unusual orientation is related to the finite system
size and a particular realization of disorder, and it was reported
previously for periodically deformed binary
glasses~\cite{Priezjev17}.    Finally, the transition from transient
clusters to formation of a system-spanning shear band of large
nonaffine displacements after a number of shear cycles was also
observed for poorly annealed, homogeneous binary
glasses~\cite{Priezjev18a}.

\vskip 0.05in


In this work, the response behavior of the void-space networks to
periodic loading is quantified by computing the pore size
distribution (PSD) functions. The calculations of the PSD functions
were performed using the open source Zeo++
software~\cite{Haranczyk17,Haranczyk12c,Haranczyk12}. The
algorithmic structure of the code can be briefly described as
follows. At the core of the implemented approach is the Voronoi
tessellation, which allows for a translation of the microstructural
information pertained to constituent atoms alongside with the
geometrical characteristics of the periodic unit cell into a
periodic graph representation of the void spaces between atoms. The
connectivity of the void network, obtained thereby, is also
computed. For each atom $i$, $i=1,...,N$, in the system having
neighbors $j=1,...,N$, the Voronoi cell is defined by the following
inequality $d(x,x_i)<d(x,x_j)$, where the distance $d(x,y)$ is the
Euclidean distance between the $x$ and $y$ points in the space. In
the Voronoi decomposition, the cells are defined as lines that are
equidistant from three neighboring atoms. The Voronoi nodes are
given by spatial positions, such that they are equidistant from four
neighboring atoms. Thereby constructed edges and nodes provide a
three-dimensional graph that represents the pores and channels. The
nodes of the graph are given by the local maxima of the function
$f(x)=\text{min}\{d(x,x_i):i=1,...,N\}$. The specifics of
implementation of the algorithm derive from a modified VORO++
software library, developed in Ref.\,\cite{Rycroft09}. As explicit
in the foregoing, the computational tool provides all the essential
information needed for complete characterization of a porous
material system.  That includes the surface areas of the pores and
pore size distribution functions. In the former case, a Monte Carlo
sampling is used for calculations, reported herein. In the present
work, the number of samples per atom was fixed at 50000. The probe
radius was chosen to be $0.3\,\sigma$. As our studies show, the
results are relatively insensitive to the probe radius, for
probe-radius magnitudes less than $0.8\,\sigma$.

\vskip 0.05in

\begin{figure}[t]
\includegraphics[width=12.cm,angle=0]{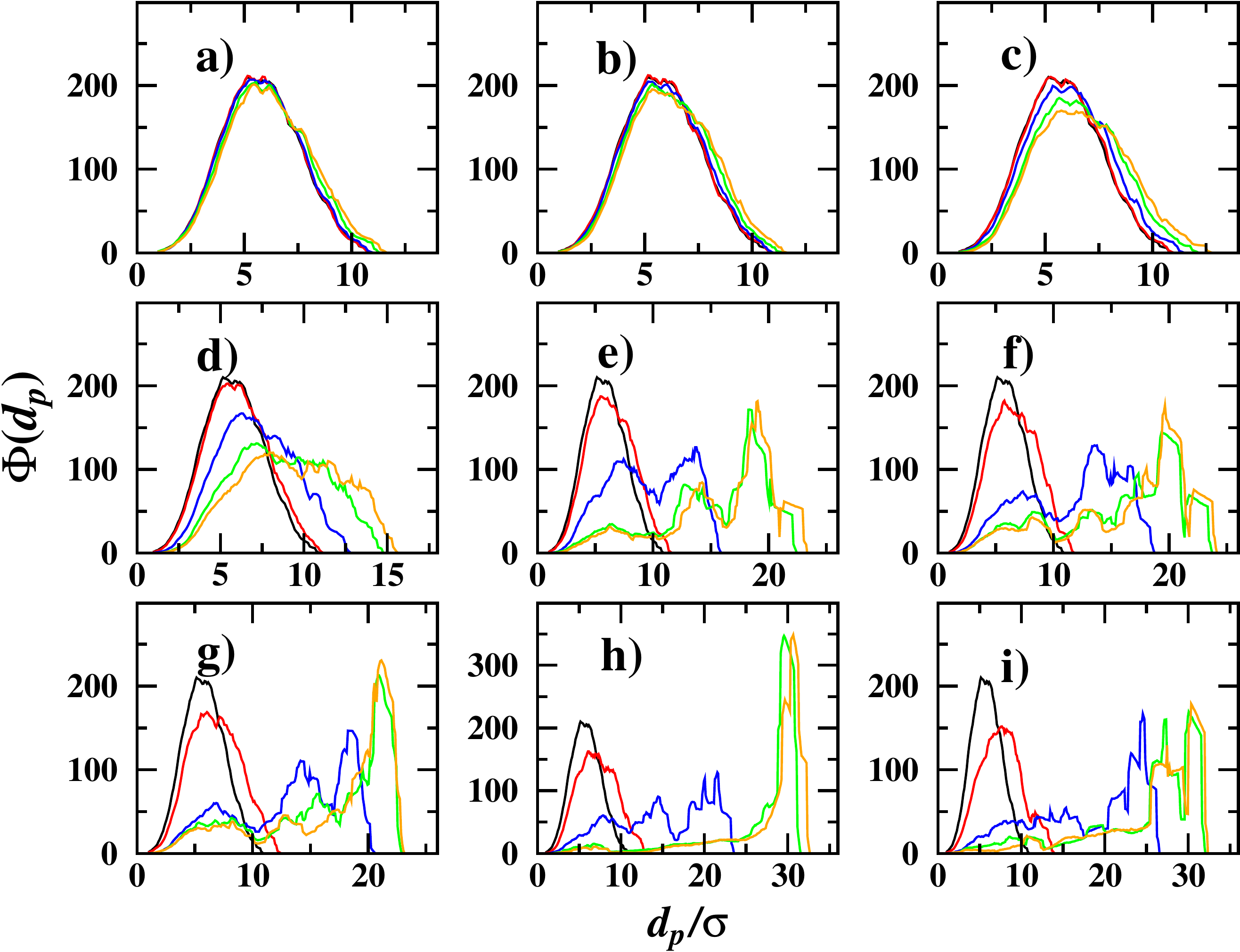}
\caption{(Color online) The distributions of pore sizes for the
strain amplitudes, $\gamma_0$: (a) 0.0, (b) 0.01, (c) 0.02, (d)
0.04, (e) 0.06, (f) 0.07, (g) 0.08, (h) 0.10, and (i) $0.12$. The
cycle number is indicated by: 0 (black), 1 (red), 10 (blue), 100
(green), and 500 (orange).  The data sets were averaged over 20
points for clarity. The black curves are the same in all panels. }
\label{fig:pore_size_strain_amp}
\end{figure}

The distributions of pore sizes during cyclic loading are reported
in Fig.\,\ref{fig:pore_size_strain_amp} for the strain amplitudes
$\gamma_0$: (a) 0.0, (b) 0.01, (c) 0.02, (d) 0.04, (e) 0.06, (f)
0.08, (g) 0.10, and (h) 0.12.  It can be observed in
Fig.\,\ref{fig:pore_size_strain_amp}\,(a-b) that in the quiescent
sample and periodically driven glass with the strain amplitude
$\gamma_0=0.01$, the shape of the distribution functions remains
nearly the same. With increasing strain amplitude below yielding
transition, the pore size distributions become slightly skewed
towards larger pore sizes, as shown in
Fig.\,\ref{fig:pore_size_strain_amp}\,(c-d).   This observation
correlates with the appearance of large-scale irreversible
displacements during the first ten cycles in
Fig.\,\ref{fig:collective_gamma0_04}, which facilitate pore
redistribution.  The influence of cyclic loading on the shape of
pore size distributions becomes significant for strain amplitudes
$\gamma_0 \geqslant 0.06$.   In particular, it can be seen that a
dominant large-size pore is developed after about 100 cycles, as
shown in Fig.\,\ref{fig:pore_size_strain_amp}\,(e-i).   This trend
is supported by visual observation of system snapshots in
Figs.\,\ref{fig:snapshot_gamma0_06}-\ref{fig:snapshot_gamma0_12}.
Notice also the appearance of a high intensity peak after 500 cycles
for $\gamma_0=0.10$ in Fig.\,\ref{fig:pore_size_strain_amp}\,(h).
The formation of a large cavity in
Fig.\,\ref{fig:snapshot_gamma0_10} is reflected in the lowest
potential energy minimum attained after 500 cycles with the strain
amplitude $\gamma_0=0.10$ (see Fig.\,\ref{fig:poten_cycle_number}).


\vskip 0.05in


As explicit in the above, a periodic mechanical loading causes not
only significant restructuring in the pore ensembles but also leads
to significant atomic rearrangements in the solid domains. The
latter effect was quantified by the corresponding studies of total
energies, which are indicative of microstructural changes
reminiscent of densification. To obtain a more quantitative
atomistic-level picture of microstructural changes, we investigated
the local density distribution in the solid domains. In the present
study, the local density of solid domains, $\langle\rho\rangle_R$,
is defined as the number of atoms located within the given radial
dimension centered on a site of the cubic lattice $L\in R^3$. An
analytical form for $\langle\rho\rangle_R$ can be obtained by
employing the following procedure. For each site, $i$, of the
lattice $\langle\rho\rangle_R$, we define a closed ball, $B_R=\{ R
\in \mathbb{R}^3, \sum_{j=1}^3 R_j^2\leqslant R_0^2 \}$, such that
$R_0=|\vec{R}_0|$ is a fixed non-zero rational number. In this case,
on-site local densities for an microcanonical ensemble consisting of
$N$ atoms can be computed as $\langle\rho\rangle_R=1/B_R \int dR^3
\delta(\vec{r}_i-\vec{R})$, where the integral is taken over $B_R$,
and $i=1, 2, ..., N$ are the atomic indexes. The quantity
$\langle\rho\rangle_R$ can be regarded as deviation of the local
density from the average density of the system with homogeneous
distribution of the solid phase.

\vskip 0.05in

\begin{figure}[t]
\includegraphics[width=12.cm,angle=0]{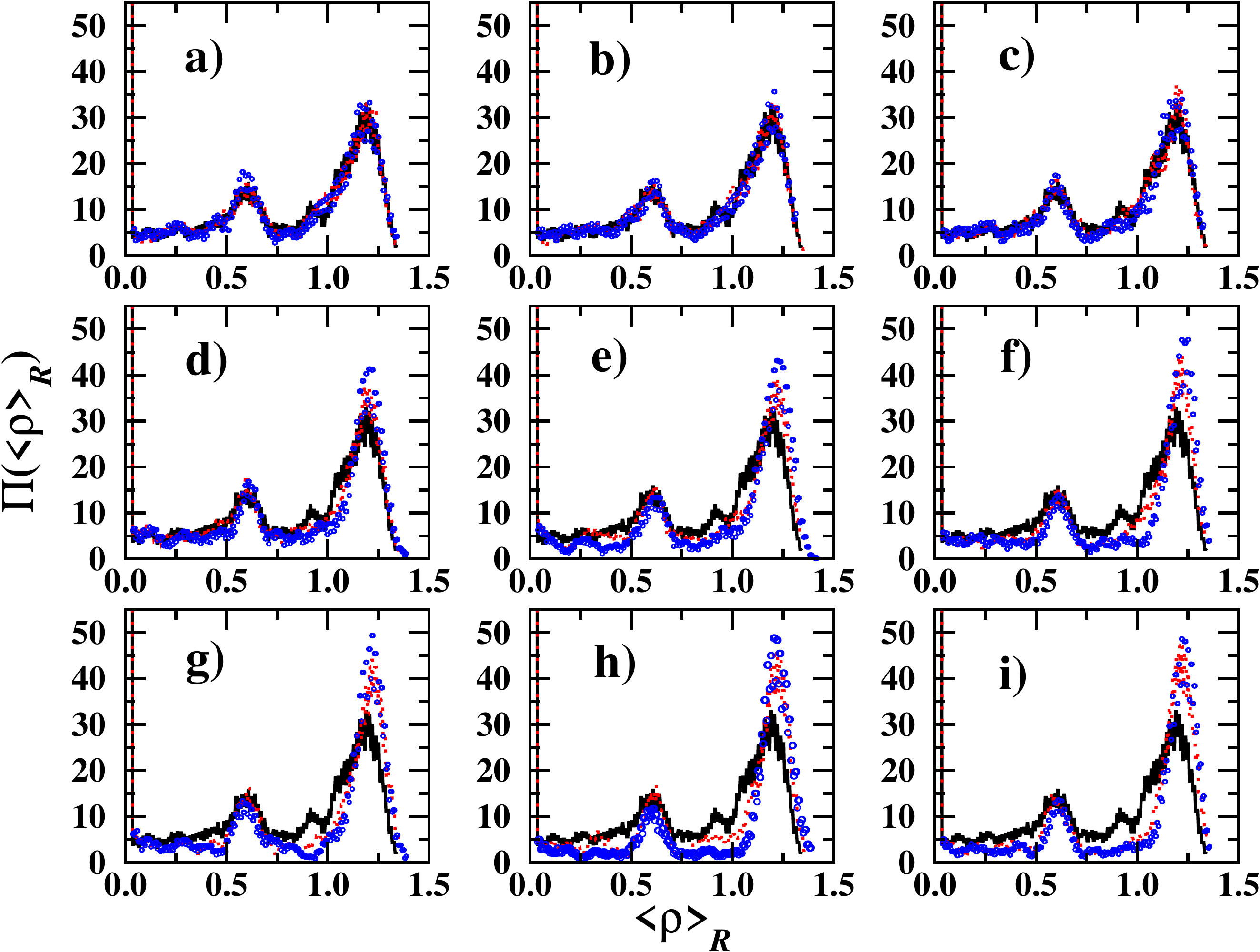}
\caption{(Color online) The local density distribution functions,
$\Pi(\langle\rho\rangle_R)$, for the strain amplitudes, $\gamma_0$:
(a) 0.0, (b) 0.01, (c) 0.02, (d) 0.04, (e) 0.06, (f) 0.07, (g) 0.08,
(h) 0.10, and (i) $0.12$.   The bin size is fixed to
$\langle\rho\rangle_R^{\text{max}}/400$ and each data set was
averaged over 20 points. In each panel, the colorcode for the cycle
number is: $0$ (solid black line), $100$ (dashed red line), and
$500$ (open blue circles). }
\label{fig:density_dist}
\end{figure}

In Fig.\,\ref{fig:density_dist}, we present the local density,
$\langle\rho\rangle_R$, distribution functions,
$\Pi(\langle\rho\rangle_R)$, computed for the porous systems under
periodic loading with the strain amplitude, $\gamma_0$: (a) 0.0, (b)
0.01, (c) 0.02, (d) 0.04, (e) 0.06, (f) 0.08, (g) 0.10, and (h)
0.12.  The main focus here is on the effect of the strain amplitude
on the structural changes taking place in the porous glassy
materials. In our previous paper~\cite{PriezMak18cyc}, we
investigated the response of porous materials to periodic loading in
the case of high porosity systems. The considered below case differs
from the previously studied system in what related to both porosity
and the form of the pore-size distribution function. The considered
herein systems have low values of porosity and, correspondingly,
Gaussian form of the pore-size distribution functions. Some of the
observations, however, are similar to those, reported in
Ref.\,\cite{PriezMak18cyc}. Thus, in general, the periodic loading
is found to cause substantial changes in density of the solid
domains. Similar to the previously considered case, the effects
differ for densities, $\rho\sigma^{3}<0.5$, and densities close to
the maximum, characteristic for the glass phase (density of
homogeneous glass). It should also be noted that the effect of the
strain amplitude is the same as the one unveiled for high-porosity
systems. Indeed, we observed that the densification in the regions
of high local density, $\rho\sigma^{3}>1.2$, is an increasing
function of the strain amplitude.  This observation holds for both
the shift of the peak position to the range of larger densities and
significant increase in the peak intensity. In the range of low
densities, $\rho\sigma^{3}<0.5$, the magnitude of the intensity
decreases. However, the intensity of the peak near
$\rho\sigma^{3}=0.6$ is approximately preserved. It should be noted
that the peak is due to the pore surface and originates from the
near-surface atoms. The above observation is indicative that the
total number of near-surface atoms remains nearly the same in the
process of pore coalescence.  This interesting observation requires
additional studies focusing on thermodynamics of pore coalescence.
The conclusions are similar to the case of lower density, considered
in Ref.\,\cite{PriezMak18cyc}. A periodic loading causes significant
homogenization of the porous glasses.  In the case considered
herein, at large strain amplitudes, dynamical evolution of the pore
structures leads to a formation of a single pore of large-size, and
significant densification.



\vskip 0.05in


In Ref.\,\cite{PriezMak18cyc}, we investigated some dynamical
aspects of the micropores evolution in porous glasses subjected to
periodic shear deformation. The systems, considered therein, were
characterized by a relatively large porosity, and the corresponding
average density of the microporous material is $\rho\sigma^{3} =
0.5$. As we have shown in Ref.\,\cite{Makeev18}, systems with large
porosity demonstrate both rather complex pore-size distributions and
highly non-trivial topology of micropores involving isolated
micropore structures and channels.  On the other hand, at higher
densities, the pore configurations primarily consist of separated by
relatively large distances micropores and their sizes are
distributed according to a Gaussian.   As was noted in
Ref.\,\cite{Makeev18}, the Gaussian ensembles of pores are
characteristic for dense glassy systems, as has been previously
observed in experimental studies~\cite{Yavari05,Dlubek96}. The
transition to the regime characterized by the Gaussian distribution
has also been discussed in detail in Ref.\,\cite{Makeev18}. One of
the key observations of the present study is the apparent
distinction between changes in the structure of micropore ensembles
in systems with high (see Ref.\,\cite{PriezMak18cyc}) and low
porosity.

\vskip 0.05in



\begin{figure}[t]
\includegraphics[width=12.cm,angle=0]{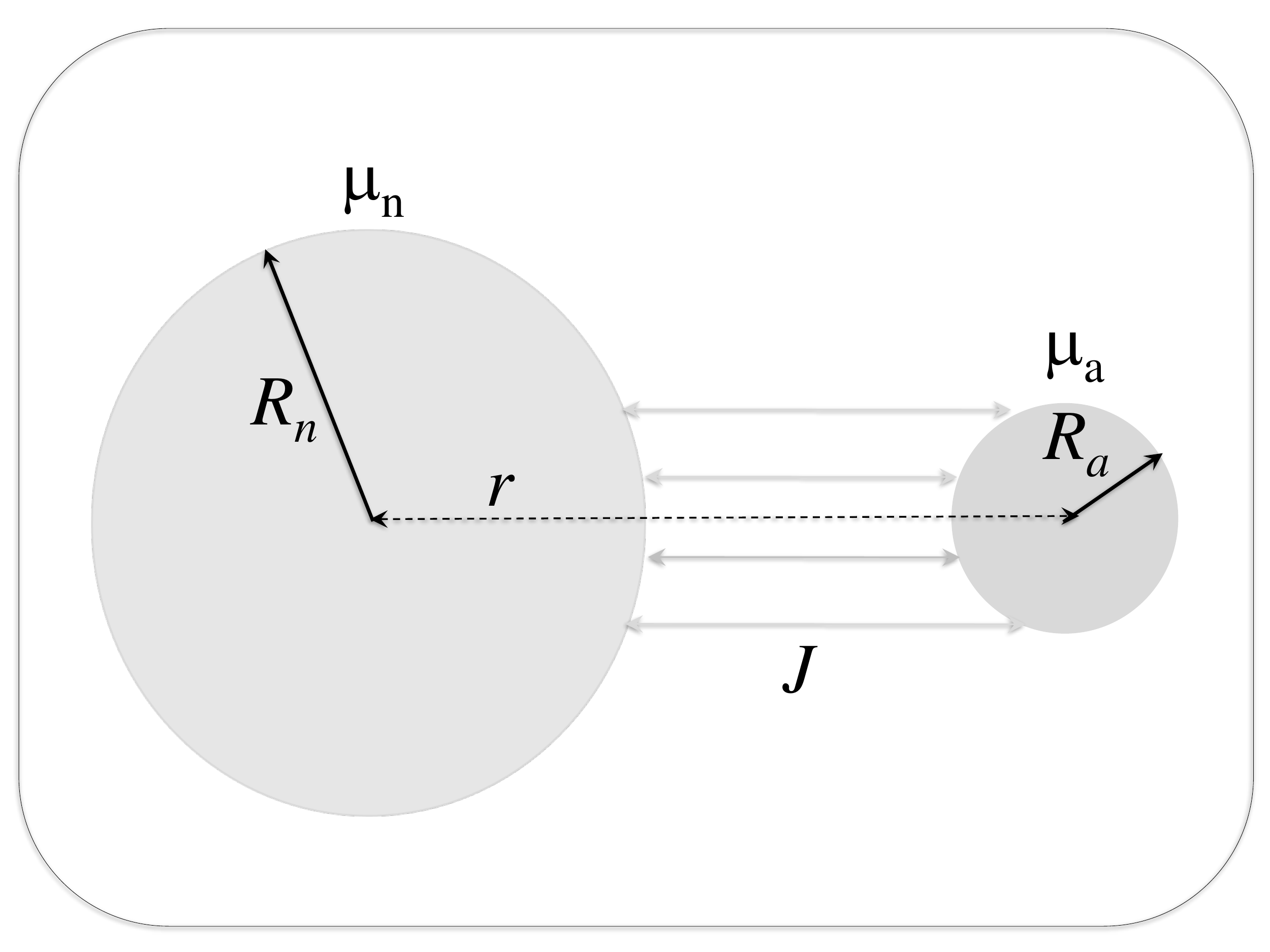}
\caption{A schematic illustration of the pore annihilation
mechanism. A small pore with radius, $R_a$, located in a vicinity of
a neighboring pore with radius, $R_n$, undergoes an instability due
to a temperature increase or an external mechanical perturbation.
Atomic rearrangements occur due to the gradient of chemical
potential, $\nabla \mu$. The structural rearrangements, leading to
the small pore annihilation, are described by the atomic current
flux into the pore, $J$, defined by the Fick's law. }
\label{fig:pore_schematic}
\end{figure}

With some restrictions, we can define a low-porosity system, as a
system that shows a Gaussian distribution of pore-sizes. In such
systems, the large amplitude loading leads to a formation of large
micropores, accompanied by annihilation of the small ones via
diffusion-driven coalescence. As we show below, this behavior can be
explained on the basis of a simple theoretical framework. To this
end, let us consider a system comprised of two pores in a glassy
medium. An illustration of the problem under consideration is given
in Fig.\,\ref{fig:pore_schematic}. Note that, in general, the
problem of pore coalescence is rather complex and involves elastic
interaction between pores, matrix material properties dependence on
porosity, and spatial position-dependent chemical potential.  In
what follows, we present a simplified theory, which, however,
describes well the underlying mechanism of micropore coalescence. As
shown in Fig.\,\ref{fig:pore_schematic}, we consider two micropores,
separated by radial distance $r$. The chemical potential of an atom
in the bulk of solid materials is defined as $\mu_{0}$ and the
chemical potential of atoms in close vicinity of the two pores are
denoted as $\mu_{a}$ and $\mu_{n}$, respectively. To approximate the
chemical potential, we employ the Kelvin's
equation~\cite{Thomson1871}:
\begin{equation}\label{eq1}
\mu = \mu_{0}+\Omega\,\sigma_s/r_{s},
\end{equation}
where $\Omega$ is the atomic volume, $\sigma_s$ is the surface
tension, and $r_s$ is the radius of pore's curvature. If the pore
closure is fully defined by mass transfer, the lifetime of the pore
is given by $\tau_{\ell} \sim S_{s}/J \sim R_{p}^{2}/J$, where
$S_{s}$ is the initial pore area and $J$ is the current into the
pore; the latter is related to the current density, $j$, and surface
area of the pore, $S_s$, by $J = j S_{s}$. In turn, the current
density, $j$, is related to the chemical potential, given by
Eq.\,(\ref{eq1}), as:
\begin{equation}\label{eq2}
j = - \frac{D_{T} c}{k_{B}T}\,\nabla\mu,
\end{equation}
where $c$ is the concentration, $T$ is temperature, and $D_{T}$ in
our case is the local strain-induced diffusion coefficient that
reflects the enhanced mobility of atoms in shear bands. Next, we
approximate Eq.\,(\ref{eq2}) by the following expression:
\begin{equation}\label{eq3}
j = \frac{D_{T} c}{ k_{B} T}\, (\mu_{a} - \mu_{n})/r,
\end{equation}
where $\mu_{a}$ is the approximation for the chemical potential of
atoms near the micropore under consideration using Eq.\,(\ref{eq1})
and $\mu_{n}$ is the approximation for the chemical potential of
atoms a neighboring micropore or sample surface (see
Fig.\,\ref{fig:pore_schematic}). Note that in both cases -- that is,
a neighboring pore and the sample surface, the chemical potential is
defined by the curvature. For some realizations of the  pore
geometries, a big pore can be regarded as nearly flat surface and
therefore the existence of sample surface can be disregarded if the
pore closure is due to a growth of a neighboring pore.  Let us omit
all the external stress fields and consider a case, when the
chemical potential is defined largely by the surface tension. The
model is supported by a number of observations, presented in
foregoing discussion.  First, the pore coalescence leads to a
decrease in potential energy and thus corresponds to an
energetically more favorable configuration (see
Fig.\,\ref{fig:poten_cycle_number}). Second, periodic deformation
induces an enhanced diffusivity in large-scale domains, as shown in
Figs.\,\ref{fig:collective_gamma0_04}--\ref{fig:collective_gamma0_10}.
This observation supports the assumption that the mass transfer into
pores occurs in regions where the pore surface intersects a shear
band.  In the following, we restrict ourselves to the classical
Kelvin's problem, Eq.\,(\ref{eq1}).  In this case, the chemical
potential is fully defined by the curvature of the pore via
Eq.\,(\ref{eq1}). Correspondingly,
\begin{equation}\label{eq4}
\mu_{a} - \mu_{n} \sim S_{s} \, \sigma_s \, (1/R_{s} - 1/R_{n}).
\end{equation}
Using Eq.\,(\ref{eq4}), we readily arrive at the following
expression for the lifetime, $\tau_{\ell}$:
\begin{equation}\label{eq5}
\tau_{\ell} \sim \frac{ k_{B} T }{D_{T} c \, \sigma_s} \, \frac{
R_{s}^{2} \, r}{1 - R_{s}/R_{n}}.
\end{equation}
The theoretical framework, developed above, derives from the
original ideas, put forth by Herring in Ref.\,\cite{Herring50}. A
number of interesting conclusions can be deduced from the above
formula.  For example, systems comprised of a pair of micropores of
the same geometry remain stable. This is not what we observe in our
simulations. The discrepancy is due to the fact that the instability
in our model systems is caused by the externally applied loading.
An extension of the model is needed to include all the relevant
stress fields.  Next observation is that the micropore lifetime is
an increasing function of its radius.  This finding is consistent
with our observation that small micropores are less stable compared
to their counterparts having larger dimensions. Also, an increase in
diffusivity makes the pore lifetime shorter, as Eq.\,(\ref{eq5})
predicts the inverse linear dependence on the diffusion coefficient.
Note that Eq.\,(\ref{eq5}) provides only a crude approximation to
the processes taking place during micropore ensembles
reorganization. It appears, however, that it can be a good starting
point for developing a deeper theoretical understanding of the
phenomenon under consideration.

\section{Conclusions}

In summary, we investigated the structural transformation and
dynamic response of microporous glassy materials to periodic shear
deformation using molecular dynamics simulations. We found that
periodic loading leads to a significant reorganization in the
structure of both void patterns and density of solid domains. The
nature of reorganization and its extent both depend strongly on the
strain amplitude and the number of loading cycles.  Moreover, the
structural rearrangements of micropores in periodically driven
systems are associated with a gradual transition to lower energy
states. The observed changes in the potential energy are attributed
to a decrease in the pore surface energy, taking place due to
micropores coalescence into voids with significantly larger
dimensions and densification of the solid domains. The two
observations are quantified by computing the pore-size and local
density distribution functions; the former accounts for the
void-spaces evolution and the latter for the density of solid
domains.

\vskip 0.05in

Furthermore, we showed that periodic deformation leads to enhanced
mobility of atoms in the whole system during the transient cycles,
which are followed by the formation of permanent shear bands at
sufficiently large strain amplitudes.    We identified and
quantified the effects of increased mobility by computing the atomic
displacements and separated the atoms into groups based upon their
displacements during each cycle.  In addition, we found that the
temporal evolution of the micropore ensembles from the initial
configurations of small-size pores into larger-scale agglomerates is
facilitated by the strain-induced diffusion of atoms in extended
domains of high mobility. Based upon assumption of strain-activated
nature of the micropore evolution, we developed an approximate
expression for the pore lifetime depending on its size and surface
energy.  Despite its simplicity, the scaling analysis correctly
captures the key findings regarding the pore coalescence deduced
from the atomistic simulations.

\section*{Acknowledgments}

Financial support from the National Science Foundation (CNS-1531923)
is gratefully acknowledged.  The article was prepared within the
framework of the Basic Research Program at the National Research
University Higher School of Economics (HSE) and supported within the
framework of a subsidy by the Russian Academic Excellence Project
`5-100'.  The molecular dynamics simulations were performed using
the LAMMPS code~\cite{Lammps}. The distributions of pore sizes were
computed using the open-source software ZEO++ developed at the
Lawrence Berkeley National
Laboratory~\cite{Haranczyk12c,Haranczyk12,Haranczyk17}.
Computational work in support of this research was performed at
Wright State University's High Performance Computing Facility and
the Ohio Supercomputer Center.


%

%

%

%

%

%

%

%


%

%

%

%

%

%

\bibliographystyle{prsty}

\begin{thebibliography}{99}






\bibitem{SchroersNat13}  B. Sarac and J. Schroers,
                         Designing tensile ductility in metallic glasses,
                         Nat. Commun. {\bf 4}, 2158  (2013).


\bibitem{Bargmann14}   B. Sarac, B. Klusemann, T. Xiao, and S. Bargmann,
                       Materials by design: An experimental and computational investigation
                       on the microanatomy arrangement of porous metallic glasses,
                       Acta Mater. {\bf 77}, 411 (2014).


\bibitem{GaoSci16}     M. Gao, J. Dong, Y. Huan, Y.~T. Wang, and W.-H. Wang,
                       Macroscopic tensile plasticity by scalarizating stress distribution
                       in bulk metallic glass,
                       Sci. Rep. {\bf 6}, 21929 (2016).

\bibitem{EckertAct16}  D. Sopu, C. Soyarslan, B. Sarac, S. Bargmann, M. Stoica, and J. Eckert,
                       Structure-property relationships in nanoporous metallic glasses,
                       Acta Mater. {\bf 106}, 199 (2016).


\bibitem{Song17}       H.~Y. Song, S. Li, Y.~G. Zhang, Q. Deng, T.~H. Xu, and Y.~L. Li,
                       Atomic simulations of plastic deformation behavior of Cu$_{50}$Zr$_{50}$ metallic glass,
                       J. Non-Cryst. Solids {\bf 471}, 312 (2017).


\bibitem{Luo18}        Y. Luo, G. Yang, Y. Shao, and K. Yao,
                       The effect of void defects on the shear band nucleation of metallic glasses,
                       Intermetallics {\bf 94}, 114 (2018).


\bibitem{ZhouCMS18}    X. Zhou, L. Wang, and C.~Q. Chen,
                       Strengthening mechanisms in nanoporous metallic glasses,
                       Comput. Mater. Sci. {\bf 155}, 151 (2018).





\bibitem{Priezjev17s}  N.~V. Priezjev and M.~A. Makeev,
                       Evolution of the pore size distribution in sheared binary glasses,
                       Phys. Rev. E {\bf 96}, 053004 (2017).


\bibitem{Priezjev18t}  N.~V. Priezjev and M.~A. Makeev,
                       Strain-induced deformation of the porous structure in binary glasses under tensile loading,
                       Comput. Mater. Sci. {\bf 150}, 134 (2018).

\bibitem{Priezjev18tp} N.~V. Priezjev and M.~A. Makeev,
                       Structural relaxation of porous glasses due to internal stresses and
                       deformation under tensile loading at constant pressure,
                       arXiv:1808.04033 (2018).

\bibitem{Priezjev18c}  N.~V. Priezjev and M.~A. Makeev,
                       Structural transformations and mechanical properties of porous glasses under compressive loading,
                       J. Non-Cryst. Solids (2018).
                       In press. DOI: https://doi.org/10.1016/j.jnoncrysol.2018.04.008




\bibitem{Priezjev13}   N.~V. Priezjev,
                       Heterogeneous relaxation dynamics in amorphous materials under cyclic loading,
                       Phys. Rev. E {\bf 87}, 052302 (2013).

\bibitem{Sastry13}     D. Fiocco, G. Foffi, and S. Sastry,
                       Oscillatory athermal quasistatic deformation of a model glass,
                       Phys. Rev. E {\bf 88}, 020301(R) (2013).

\bibitem{Reichhardt13} I. Regev, T. Lookman, and C. Reichhardt,
                       Onset of irreversibility and chaos in amorphous solids under periodic shear,
                       Phys. Rev. E {\bf 88}, 062401 (2013).

\bibitem{Priezjev14}   N.~V. Priezjev,
                       Dynamical heterogeneity in periodically deformed polymer glasses,
                       Phys. Rev. E {\bf 89}, 012601 (2014).

\bibitem{IdoNature15}  I. Regev, J. Weber, C. Reichhardt, K.~A. Dahmen, and T. Lookman,
                       Reversibility and criticality in amorphous solids,
                       Nat. Commun. {\bf 6}, 8805 (2015).

\bibitem{Priezjev16}   N.~V. Priezjev,
                       Reversible plastic events during oscillatory deformation of amorphous solids,
                       Phys. Rev. E {\bf 93}, 013001 (2016).


\bibitem{Kawasaki16}   T. Kawasaki and L. Berthier,
                       Macroscopic yielding in jammed solids is accompanied by a non-equilibrium
                       first-order transition in particle trajectories,
                       Phys. Rev. E {\bf 94}, 022615 (2016).


\bibitem{Yang16}       Y.~F. Ye, S. Wang, J. Fan, C.~T. Liu, and Y. Yang,
                       Atomistic mechanism of elastic softening in metallic glass under cyclic
                       loading revealed by molecular dynamics simulations,
                       Intermetallics {\bf 68}, 5 (2016).

\bibitem{Priezjev16a}  N.~V. Priezjev,
                       Nonaffine rearrangements of atoms in deformed and quiescent binary glasses,
                       Phys. Rev. E {\bf 94}, 023004 (2016).


\bibitem{Sastry17}     P. Leishangthem, A.~D.~S. Parmar, and S. Sastry,
                       The yielding transition in amorphous solids under oscillatory shear deformation,
                       Nat. Commun. {\bf 8}, 14653 (2017).


\bibitem{Priezjev17}   N.~V. Priezjev,
                       Collective nonaffine displacements in amorphous materials during large-amplitude oscillatory shear,
                       Phys. Rev. E {\bf 95}, 023002 (2017).

\bibitem{Hecke17}      S. Dagois-Bohy, E. Somfai, B.~P. Tighe, and M. van Hecke,
                       Softening and yielding of soft glassy materials,
                       Soft Matter {\bf 13}, 9036 (2017).

\bibitem{Keblinsk17}   R. Ranganathan, Y. Shi, and P. Keblinski,
                       Commonalities in frequency-dependent viscoelastic damping in glasses
                       in the MHz to THz regime,
                       J. Appl. Phys. {\bf 122}, 145103 (2017).

\bibitem{Priezjev18}   N.~V. Priezjev,
                       Molecular dynamics simulations of the mechanical annealing process in
                       metallic glasses: Effects of strain amplitude and temperature,
                       J. Non-Cryst. Solids {\bf 479}, 42 (2018).

\bibitem{Priezjev18a}      N.~V. Priezjev,
                           The yielding transition in periodically sheared binary glasses at finite temperature,
                           Comput. Mater. Sci. {\bf 150}, 162 (2018).


\bibitem{NVP18strload}     N.~V. Priezjev,
                           Slow relaxation dynamics in binary glasses during stress-controlled,
                           tension-compression cyclic loading,
                           Comput. Mater. Sci. {\bf 153}, 235 (2018).


\bibitem{SastryBands18}    A.~D.~S. Parmar, S. Kumar, and S. Sastry,
                           Strain localisation above the yielding point in cyclically deformed glasses,
                           arXiv:1806.02464 (2018).


\bibitem{PriezMak18cyc}    N.~V. Priezjev and M.~A. Makeev,
                           The influence of periodic shear on structural relaxation and pore redistribution in binary glasses,
                           arXiv:1808.09323 (2018).





\bibitem{KobAnd95}     W. Kob and H.~C. Andersen,
                       Testing mode-coupling theory for a supercooled binary Lennard-Jones mixture:
                       The van Hove correlation function,
                       Phys. Rev. E {\bf 51}, 4626 (1995).


\bibitem{Allen87}      M.~P. Allen and D.~J. Tildesley,
                       {\it Computer Simulation of Liquids} (Clarendon, Oxford, 1987).


\bibitem{Lammps}       S.~J. Plimpton,
                       Fast parallel algorithms for short-range molecular dynamics,
                       J. Comp. Phys. {\bf 117}, 1 (1995).


\bibitem{Kob11}        V. Testard, L. Berthier, and W. Kob,
                       Influence of the glass transition on the liquid-gas spinodal decomposition,
                       Phys. Rev. Lett. {\bf 106}, 125702 (2011).

\bibitem{Kob14}        V. Testard, L. Berthier, and W. Kob,
                       Intermittent dynamics and logarithmic domain growth during
                       the spinodal decomposition of a glass-forming liquid,
                       J. Chem. Phys. {\bf 140}, 164502 (2014).

\bibitem{Makeev18}     M.~A. Makeev and N.~V. Priezjev,
                       Distributions of pore sizes and atomic densities in binary mixtures
                       revealed by molecular dynamics simulations,
                       Phys. Rev. E {\bf 97}, 023002 (2018).


\bibitem{Haranczyk17}  D. Ongari, P.~G. Boyd, S. Barthel, M. Witman, M. Haranczyk, and B. Smit,
                       Accurate characterization of the pore volume in microporous crystalline materials,
                       Langmuir {\bf 33}, 14529 (2017).

\bibitem{Haranczyk12c} R.~L. Martin, B. Smit, and M. Haranczyk,
                       Addressing challenges of identifying geometrically diverse sets of crystalline porous materials,
                       J. Chem. Inf. Model. {\bf 52}, 308 (2012).

\bibitem{Haranczyk12}  T.~F. Willems, C.~H. Rycroft, M. Kazi, J.~C. Meza, and M. Haranczyk,
                       Algorithms and tools for high-throughput geometry-based analysis of crystalline porous materials,
                       Micropor. Mesopor. Mater. {\bf 149}, 134 (2012).

\bibitem{Rycroft09}    C.~H. Rycroft,
                       VORO++: A three-dimensional Voronoi cell library in C++,
                       Chaos {\bf 19}, 041111 (2009).

\bibitem{Yavari05}     A. R. Yavari, A. Le Moulec, A. Inoue, N. Nishiyama, N. Lupu,
                       E. Matsubara, W. J. Botta, G. Vaughan, M. Di Michiel, and A. Kvick,
                       Excess free volume in metallic glasses measured by X-ray diffraction,
                       Acta Mater. {\bf 53}, 1611 (2005).

\bibitem{Dlubek96}     G. Dlubek, A. P. Clarke, H. M. Fretwell, S. B. Dugdale, and M. A. Alam,
                       Positron lifetime studies of free volume hole size
                       distribution in glassy polycarbonate and polystyrene,
                       Phys. Stat. Sol. (A) {\bf 157}, 351 (1996).


\bibitem{Thomson1871}  W. Thomson,
                       On the equilibrium of vapour at a curved surface of liquid,
                       Philosophical Magazine {\bf 42}, 448-452 (1871).

\bibitem{Herring50}    C. Herring,
                       Effect of change of scale on sintering phenomena,
                       J. Appl. Phys. {\bf 21}, 301 (1950).




\end{thebibliography}

\end{document}